\documentclass{ptapap}
\usepackage{amsmath}
\usepackage{amssymb}
\usepackage{float}
\usepackage{gensymb}

\author{Praneet Bhoj}[UCSD]
\author{Roman Gerasimov}[UCSD]
\author{Brian Keating}[UCSD]
\affil[UCSD]{Center for Astrophysics and Space Sciences, University of California, San Diego}

\title{What is Known About the Polarization of Starlight in the Southern Hole}

\begin{document}
\maketitle

\begin{abstract}
Among the greatest mysteries in cosmology are the flatness problem, concerning the lack of curvature of the universe, and the homogeneity problem, questioning why the universe is almost isotropic despite having regions that are causally disconnected. These problems served as motivation for the theory of inflation, which suggests a period of exponential expansion in the early universe, and the inflationary origin of the universe can be traced by B-mode polarization. In an effort to better understand the potential foreground systematics, especially the levels of polarized dust emission, we queried the Heiles catalog to produce a list of starlight polarization data in the so-called ``Southern Hole'', which is an approximately $20\times20$ degree region centered at RA: $00\textsuperscript{h}12\textsuperscript{m}00\textsuperscript{s}$ and DEC: $-59\degree18'00''$ that is being examined by multiple CMB polarization experiments. Because magnetic field tends to dictate the orientation of dust grains, which in turn determines how starlight is polarized, starlight polarization can be used to trace magnetic fields. Therefore, to improve our understanding of the properties of this region, we used this catalog, along with Gaia data as tracers of the three-dimensional distribution of dust, as a potential indicator of magnetic field orientation throughout the galaxy in the Southern Hole region. We then analyzed these data with the hope that magnetic field data can be used to create a template to aid in subtracting away the contamination of CMB B-mode searches by polarized dust emission. While the results of the analysis are promising, we found that the currently available data are severely inadequate for the purpose of creating a template, thus demonstrating the need for improved and more uniform coverage of the Southern Hole when it comes to polarization measurements.
\end{abstract}

\section{Introduction}
The cosmic microwave background (CMB) is electromagnetic radiation left over from the early universe that produces a nearly uniform signal from every direction in space. The CMB has been measured for many different experiments, but it is especially important for experiments seeking to support the idea of inflation, which is a theory that suggests the early universe underwent a period of exponential expansion. The theory predicts that inflation would cause gravitational waves that polarize the CMB. Because these gravitational waves are the only potential source of B-mode polarization, searching for CMB B-modes is one of the primary ways to provide evidence for these gravitational waves and consequently the theory of inflation. There are numerous current and planned projects that seek to explore the polarization of the CMB and its implications on the theory of inflation. The Southern Hole, which is an approximately $20\times20$ degree region centered at RA: $00\textsuperscript{h}12\textsuperscript{m}00\textsuperscript{s}$ and DEC: $-59\degree18'00''$, is a region surveyed by many of these projects, including BICEP\footnote{https://www.cfa.harvard.edu/CMB/bicep2/}, WMAP\footnote{https://map.gsfc.nasa.gov/}, POLARBEAR\footnote{http://bolo.berkeley.edu/polarbear/}, Planck\footnote{https://www.cosmos.esa.int/web/planck}, South Pole Telescope\footnote{https://pole.uchicago.edu/}, and the Simons Array/Simons Observatory\footnote{https://simonsobservatory.org/}. For reference, the Southern Hole contains the constellation Tucana and partially contains the constellation Phoenix.
\newline\newline
Although the Southern Hole was chosen because of the reduced amount of dust at high galactic latitudes, there still is some dust in the region, and proper analysis of CMB polarization in this region requires the removal of the effects of those foreground dust emissions. Therefore, creating a template for the removal of those effects can be very helpful. A key concept that can be used in producing this template is the fact that starlight is polarized \citep{first_pol}, and the primary source of this polarization is dust. Because the dust grains polarize starlight based on their orientation and the orientation of dust grains is dictated by magnetic fields \citep{DG51}, analyzing starlight polarization offers insights regarding the orientation of magnetic fields that can be used to create the desired template. Our work outlines how this can be approached using existing starlight polarization data along with stellar distance data from the Gaia satellite. This aligns with the planned work of the PASIPHAE\footnote{http://pasiphae.science/} project, which states its goal is to “measure the linear polarization from millions of stars, and use these to create a three-dimensional tomographic map of the magnetic field threading dust clouds within the Milky Way” \citep{pasiphae}. Our work aims to show that starlight in the Southern Hole is polarized, polarization varies with celestial coordinates and distances, nearby stars carry similar polarization, and there is a hint of correlation between starlight polarization and Planck emission measurements. In doing so, we hope to highlight the need for PASIPHAE to cover the Southern Hole, preferably as soon as possible. This way, a number of near-term CMB experiments observing in the Southern Hole can benefit from the detailed template for removing contamination due to foreground dust emission.

\section{Methods}
The data from PASIPHAE will be very helpful for the type of analysis done in this paper, however the methods described in this paper differ from those that will be used by PASIPHAE. While our work utilizes existing starlight polarization data for analysis in the form of 2D reconstruction,
PASIPHAE plans to do a much more detailed, ``layer-by-layer'' 3D tomography by surveying regions with galactic latitudes greater than $50$ degrees and less than $-50$ degrees and measuring starlight polarization for sources within those areas. It will achieve high sensitivity of measurements by averaging polarization measurements over a given region \citep{pasiphae}.
\newline\newline
Each source with polarization data is cross-referenced with the Gaia satellite's data release \citep{gaia2018, gaia2016}, for the purpose of finding its distance measurement. Using the distance and polarization data, a three-dimensional line of sight can be constructed, with starlight polarization varying as a function of distance. The change in polarization over distance is presumably due to the presence of dust in between sources, which contributes to the polarization of starlight in different ways. Assuming that dust grains are asymmetric (they have ``short'' and ``long axes''), the grains have a tendency to align their short axes with the magnetic field \citep{andersson} and scatter light along their long axes. Therefore, starlight at optical wavelengths is polarized along the short axis (parallel to the magnetic field) while microwave dust emission is polarized along the long axis (perpendicular to the magnetic field). By analyzing how the starlight polarization angle changes between two sources along a given line of sight, predictions can be made regarding the direction of the magnetic field within that space between the sources. Repeating this over many sources and many lines of sight will potentially allow for the construction of the magnetic field template and, more importantly, a template of predicted polarized dust emission in the region.
\newline\newline
General correlation in polarization between starlight and microwave emission was demonstrated at high confidence in \cite{planck_starlight}, where publicly available optical starlight polarimetry is compared against Planck sub-millimeter measurements along the same lines of sight. Two coefficients of proportionality were derived: $R_{S/V}$ between the emission polarization fraction and the starlight polarization fraction normalized by optical depth; and $R_{P/p}$ between the emission polarization intensity ($\sqrt{Q^2+U^2}$, where $Q$ and $U$ are the Stokes parameters) and the unnormalized starlight polarization fraction. The correlations were established primarily based on high signal-to-noise polarization measurements around the galactic plane. Here, we search for a similar effect in the Southern hole and compare the scatter within it to the best fit from \cite{planck_starlight}. We specifically focus on $R_{P/p}$, as $R_{S/V}$ requires optical depth data that are not available for stars in the Southern hole with adequate polarimetric signal-to-noise-ratios in the optical depth catalogues considered in \cite{planck_starlight} (\cite{tau_1}, \cite{tau_2}, \cite{tau_3}, \cite{tau_4}).

\section{Comments on the catalogs}
\subsection{Agglomeration of stellar polarization catalogs (Heiles)}
\begin{figure}[H]
    \begin{center}
        \vspace*{-3mm}
        \hspace*{-8mm}
        \includegraphics[scale=0.3]{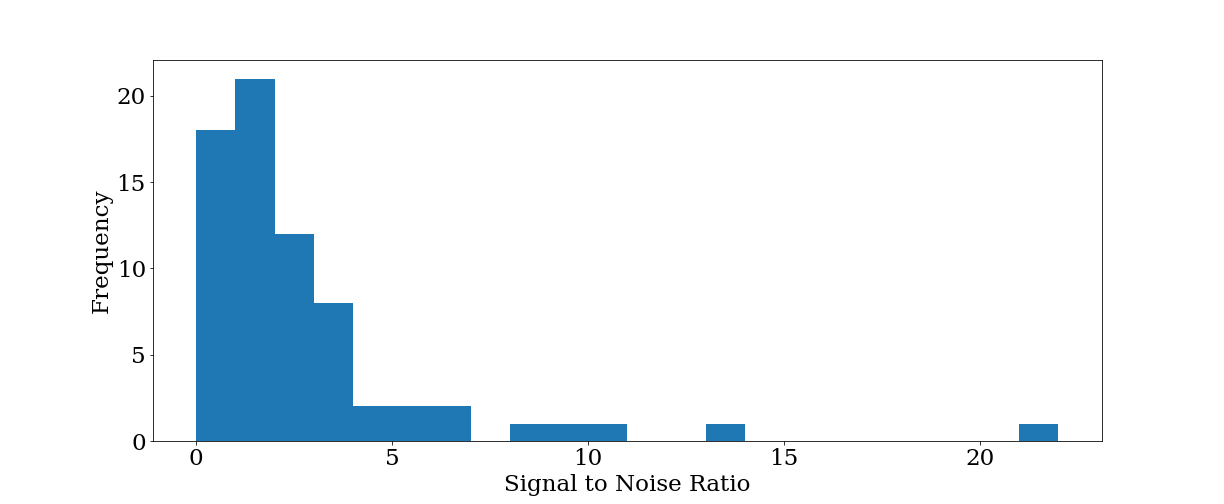}
        \vspace*{-5mm}
        \caption{Histogram of the signal-to-noise ratios for the 70 sources in the Southern Hole that are taken from the Heiles catalog.}
        \label{snr_hist}
    \end{center}
\end{figure}

The Heiles catalog \citep{heiles2000} is a compilation of stellar polarization data from various other catalogs. In total, the Heiles catalog contains various measurements and star identification information for the J2000 epoch for 9,286 different sources across the entire sky. For our work, the particularly useful pieces of information for each source in this catalog are the measurements of right ascension, declination, percent polarization, polarization angle, and magnitude. However, the catalog also has a couple downsides in the context of our work. For one thing, the catalog may not have the most accurate position and starlight polarization measurements for some of the sources. Additionally, the catalog doesn't have strong coverage in the Southern Hole region. In fact, the catalog only contains data for 70 sources within the Southern Hole (0.175 stars per square degree). Furthermore, many of the sources have a low signal-to-noise ratio. Out of the 70 Heiles sources in the Southern Hole, 18 sources have a signal-to-noise ratio below 1. The distribution of signal-to-noise ratios of the sources in the Southern Hole can be seen in Fig.~\ref{snr_hist}.

\subsection{Gaia DR2 (Gaia Collaboration)}
The Gaia DR2 catalog \citep{gaia2018, gaia2016} is the contents of the 2018 data release from the European Space Agency's Gaia\footnote{https://www.cosmos.esa.int/web/gaia} satellite. The catalog contains data for the J2000 epoch for nearly 1.7 billion sources. For our work, the relevant measurements for each source in this catalog include the right ascension, declination, stellar parallax, and magnitude.

\subsection{Microwave emission}
\label{sec:microwave_emission}
For the starlight-emission correlation analysis, we used the all-sky $353\ \mathrm{GHz}$ map from the 2018 data release of the Planck satellite \citep{planck_2018}. For each line of sight, the emission Stokes parameters were calculated as the average across all \texttt{HEALPIX} pixels that fall within $0.1$ degrees of the desired coordinates. The calculations were carried out after the map was downsampled to \texttt{NSIDE=512} (see \citep{healpix}).

\section{Cross-referencing the catalogs}
We started by using VizieR\footnote{https://vizier.u-strasbg.fr/viz-bin/VizieR-2} to query the Heiles catalog and retrieve the data of the sources positioned in the Southern Hole, thus producing a list of stars in the Southern Hole with polarization measurements. In addition to polarization data, our work involved the distances of sources as well, so those same sources with polarization measurements were searched for in the Gaia DR2 catalog. For each Heiles source in the Southern Hole, VizieR was used to search the Gaia catalog for sources within a one arcminute radius of its right ascension and declination. In doing so, it was apparent that most Heiles sources had exactly one match in the Gaia catalog within the search radius, which made it easy to identify it as the same source. However, there were some Heiles sources that had multiple matches in the Gaia catalog within that search radius. These cases were handled by selecting the Gaia match that had the magnitude measurement closest to the Heiles source. While processing each matched source, the source's distance was calculated using the parallax measurement from the Gaia catalog. 
\newline\newline
The matches for all the Heiles sources in the Southern Hole were compiled into a dataframe that then included, for each source, measurements of right ascension, declination, distance, percent polarization, polarization angle, and magnitude. Any source that didn't have a match or that had a match with a magnitude difference beyond $\pm 2$ (in which case we couldn't confidently say that the Heiles source and Gaia source were the same source) was ``flagged'' in the dataframe. There were three such sources that were flagged. The 67 non-flagged sources in the dataframe were used for all subsequent analysis.

\section{Visualizing the Southern Hole}
\begin{figure}[H]
    \begin{center}
        \vspace*{-3mm}
        \hspace*{-5mm}
        \includegraphics[scale=0.3]{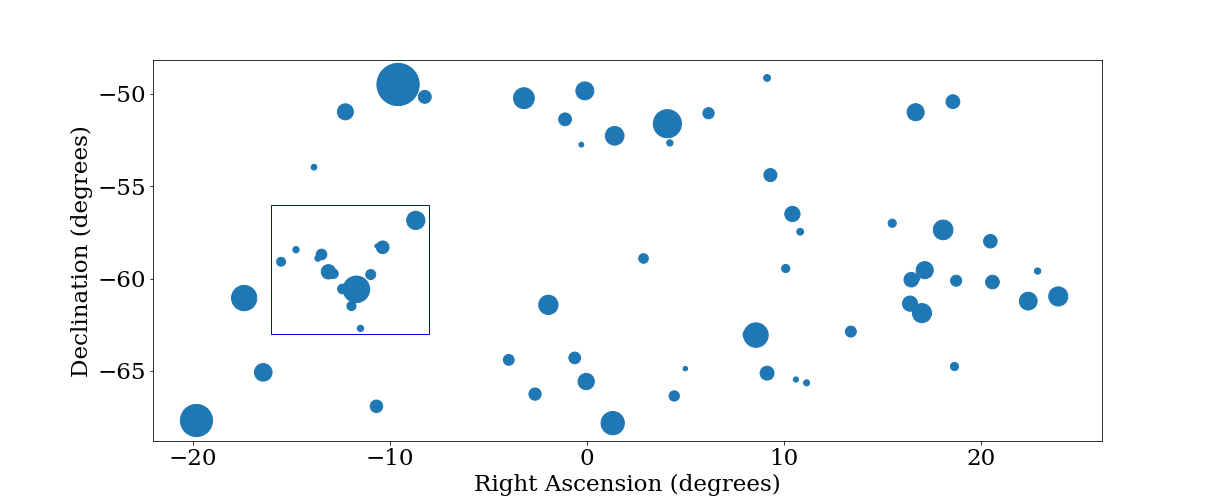}
        \vspace*{-5mm}
        \caption{Two-dimensional plot of the sources in the Southern Hole, with markers sized by percent polarization. The sources in the blue box were used in creating Fig.~\ref{pol_vs_dist} and Fig.~\ref{pa_vs_dist}.}
        \label{patch}
    \end{center}
\end{figure}

\begin{figure}[H]
    \begin{center}
        \vspace*{-6mm}
        \hspace*{-8mm}
        \includegraphics[scale=0.3]{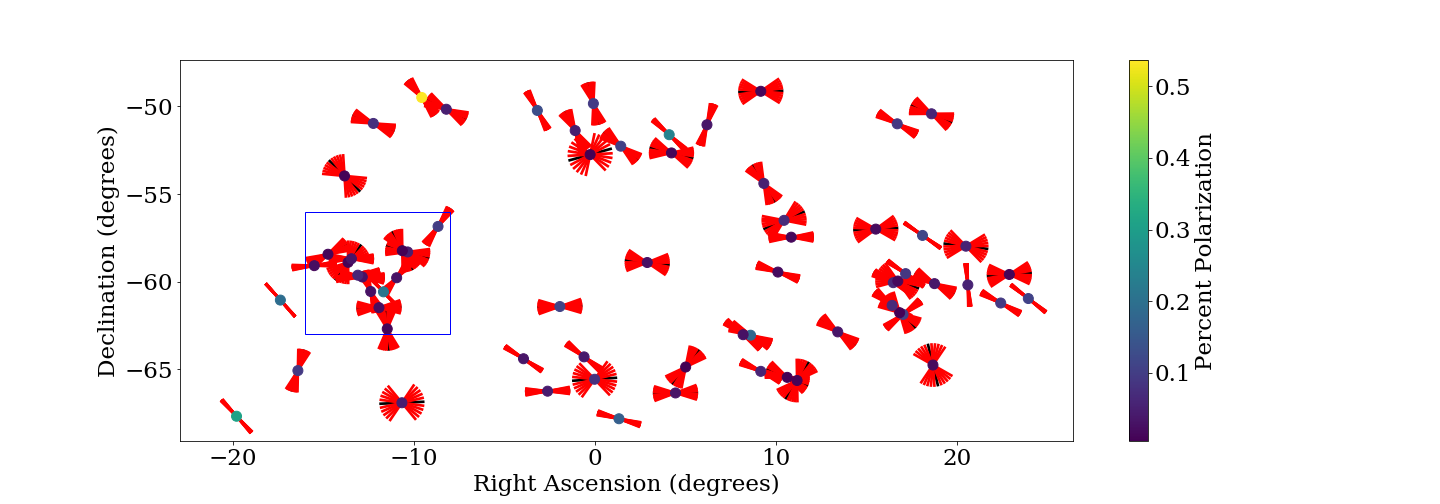}
        \vspace*{-9mm}
        \caption{Two-dimensional plot of the sources in the Southern Hole and their polarization angles, color coded by percent polarization. The sources in the blue box were used in creating Fig.~\ref{pol_vs_dist} and Fig.~\ref{pa_vs_dist}.}
        \label{patch_pa}
    \end{center}
\end{figure}

All the matched sources in the dataframe were plotted in a two-dimensional scatter plot of right ascension versus declination. Doing so revealed the distribution of sources in the Southern Hole. The data points were sized based on the percent polarization of the sources, and Fig.~\ref{patch} was produced.
\noindent In addition to this scatter plot, a ``quiver'' plot was also produced on the same axes as before to visualize the polarization angles of the sources in the Southern Hole and look for similarities/trends. The plot combines the polarization angles as well as the uncertainty in the angles to produce the ``fan'' shapes as seen in Fig.~\ref{patch_pa}.
\noindent Creating these plots of the Southern Hole made it easier to identify clusters of sources that were of interest for 3D analysis using lines of sight.

\section{Creating lines of sight}
\begin{figure}[H]
    \begin{center}
        \vspace*{-6mm}
        \hspace*{-8mm}
        \includegraphics[scale = 0.3]{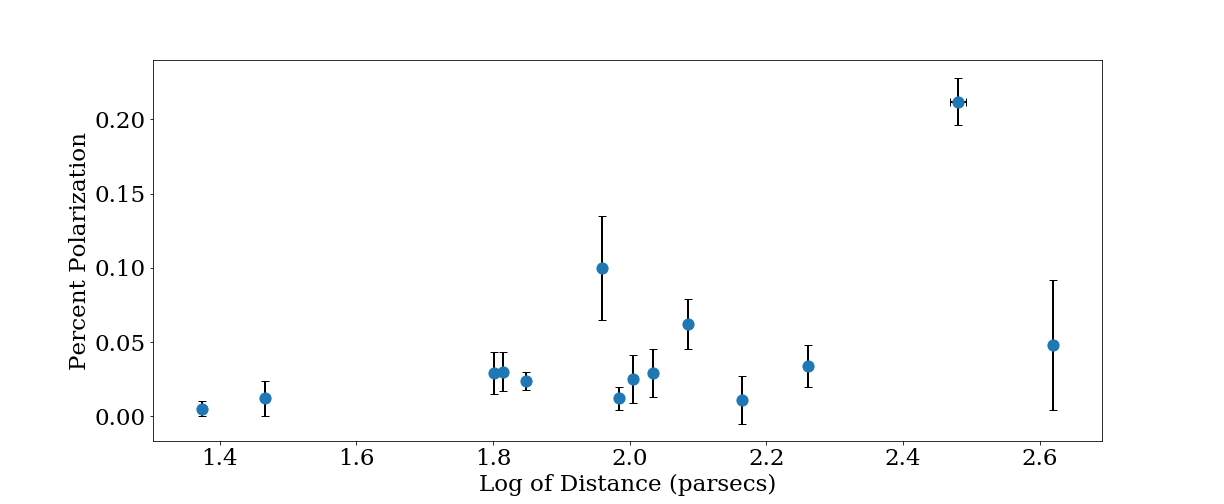}
        \vspace*{-5mm}
        \caption{Polarization versus base 10 logarithm of distance for a given cluster of target sources located within the Southern Hole. Note: all the points have polarization (vertical) and distance (horizontal) error bars, however all the sources except the source at about 300 parsecs have very small distance errors relative to the distance magnitude and therefore most of the horizontal error bars are behind the points and cannot be seen.}
        \label{pol_vs_dist}
    \end{center}
\end{figure}

\begin{figure}[H]
    \begin{center}
        \vspace*{-3mm}
        \hspace*{-8mm}
        \includegraphics[scale = 0.3]{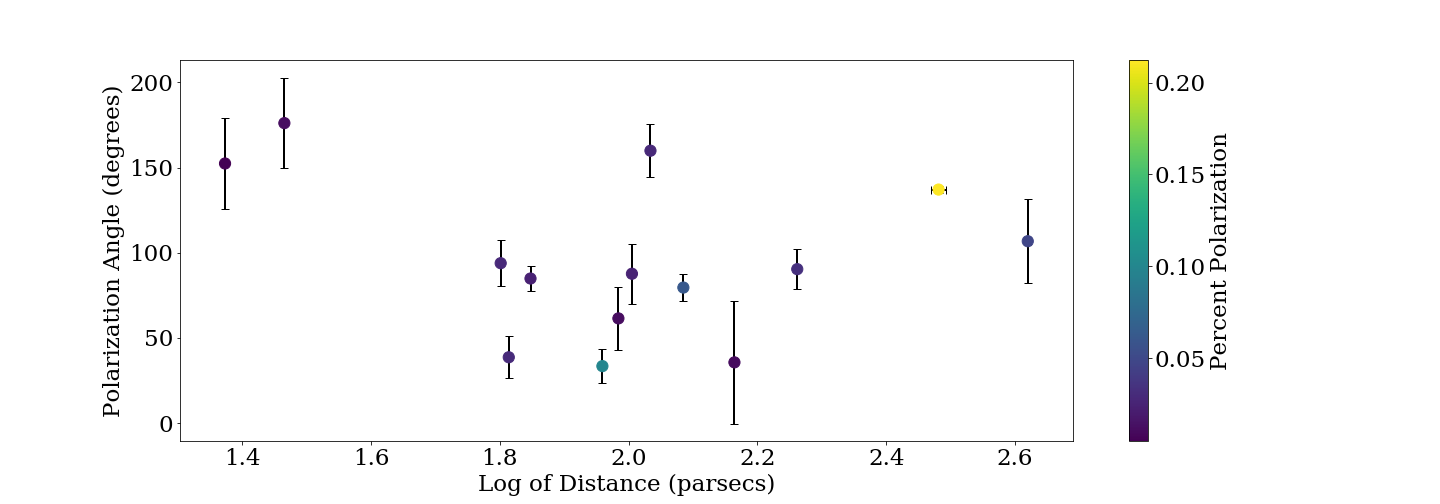}
        \vspace*{-9mm}
        \caption{Polarization angle versus the base 10 logarithm of distance for a given cluster of target sources located within the Southern Hole, color coded by percent polarization. Note: all the points have polarization angle (vertical) and distance (horizontal) error bars, however all the sources except the source at about 300 parsecs have very small distance errors relative to the distance magnitude and therefore most of the horizontal error bars are behind the points and cannot be seen.}
        \label{pa_vs_dist}
    \end{center}
\end{figure}

To analyze apparent clusters\footnote{For reference, ``clusters'' are defined here as groups of sources that are close to each other when considering only right ascension and declination (they may or may not be close to each other in terms of actual three-dimensional distance).} of sources in three dimensions, we developed a function that creates three-dimensional lines of sight for stars in a given cluster. It takes in two ordered pairs as parameters, and these coordinates define the upper left and bottom right corners of a rectangular region in the two-dimensional right ascension versus declination plot. Sources within the given rectangle are considered part of the cluster. Once the rectangle is defined, the function searches through the dataframe of sources and their corresponding positions to find which sources are within the rectangle. Those sources are referred to as ``target'' sources. From there, the coordinates of the target sources are retrieved, then the function color maps the target sources using their polarization percentages and displays their positions in a three-dimensional plot. After creating the three-dimensional plot, allowing for visualization of the line of sight, the function generates two more plots for analysis of the line of sight. Fig.~\ref{pol_vs_dist} is a polarization versus distance plot created using the cluster of sources in rectangle with an upper left corner at (RA, DEC) = (-16, -56) and bottom right corner at (RA, DEC) = (-8, -63). For reference, this region is boxed in blue in Fig.~\ref{patch} and Fig.~\ref{patch_pa}.
\noindent The other plot the function produces is a plot of polarization angle versus distance for the target sources. Fig.~\ref{pa_vs_dist} was created for the same cluster of sources as Fig.~\ref{pol_vs_dist}.
\noindent By creating these plots, we are able to examine how polarization angle varies as a function of distance within the given line of sight. The way polarization angle changes with distance is significant because if polarization angle changes between two given sources, it suggests the presence of dust clouds in between. Those clouds polarize starlight and emit polarized microwaves that can contribute to the contamination of B-mode searches. As explained earlier, the behavior of the dust grains making up those dust clouds leads to a relationship between starlight polarization angle and magnetic field orientation. Because of that, these plots can help to determine the nature dust clouds and the magnetic field direction at certain distances along the line of sight. For this line of sight, Fig.~\ref{pa_vs_dist} seems to show that the polarization angle is stable for the first couple sources, then greatly destabilizes after that. Additionally, Fig.~\ref{pol_vs_dist} appears to display a trend of increasing percent polarization along most of the line of sight, and then a drop in polarization at the end. These two trends suggest the presence of multiple dust clouds along this particular line of sight, and if these data were good enough, this would be the conclusion to work with and analyze further. However, the number of measured sources along this line of sight, and in the Southern Hole in general, is severely inadequate, and those with polarization measurements tend to have suboptimal signal to noise ratios. Therefore, the conclusions made from the plots above are far from definitive. With the additional data that the PASIPHAE project can provide by surveying the Southern Hole, the trends in these plots will become much more apparent make it possible for conclusions with more confidence. Having access to more data and performing this analysis over multiple clusters within the Southern Hole could be a big step toward the ultimate goal of creating a template of polarized dust emission.

\section{Starlight-emission correlation}

\begin{figure}[H]
    \begin{center}
        \includegraphics[scale=0.3]{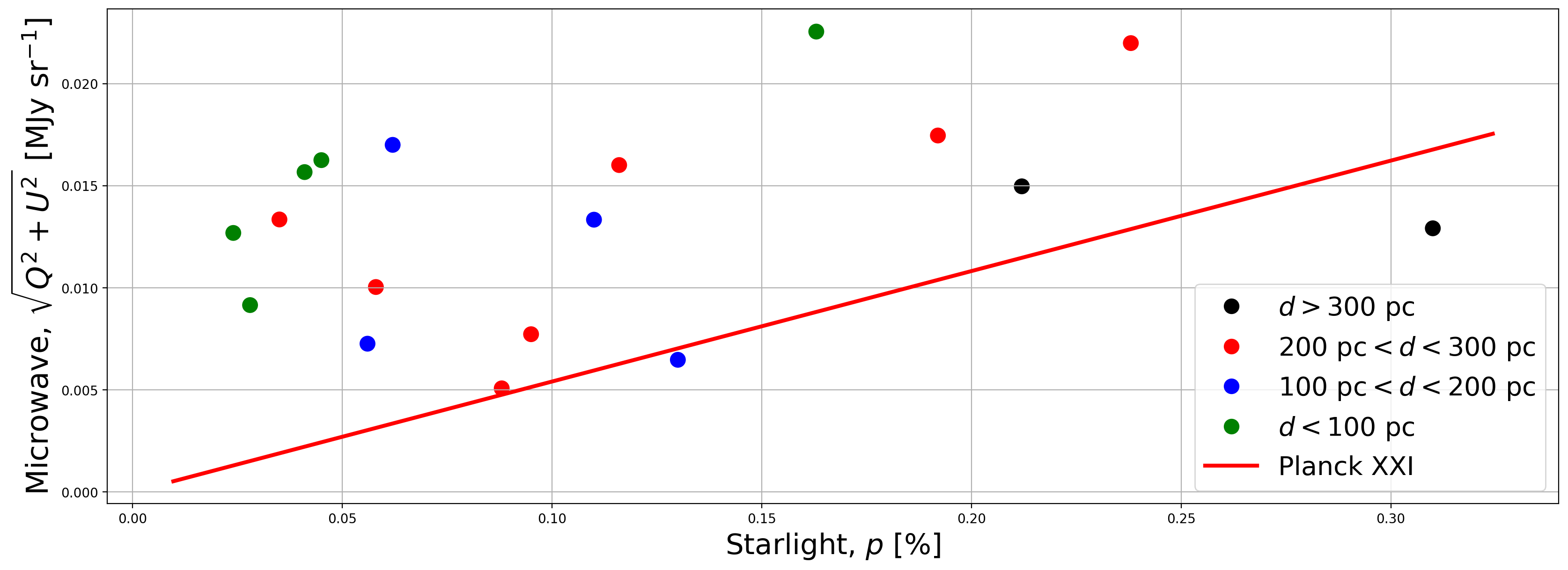}
        \vspace*{-9mm}
        \caption{Correlation between starlight polarization and microwave polarization intensity along the same lines of sight for the stars in the Southern Hole. A line of best fit is provided as estimated in \citet{planck_starlight} for a larger sample of stars.}
        \label{microwave_fig}
    \end{center}
\end{figure}

Fig.~\ref{microwave_fig} shows the emission-starlight correlation of polarization in the Southern Hole, calculated for Heiles stars within the patch, whose optical polarization was measured at the signal-to-noise ratio of $3$ or above ($18$ stars are shown in total). The microwave polarization intensity, $\sqrt{Q^2 + U^2}$, was calculated for the same lines of sight as described in section \ref{sec:microwave_emission}. The best fit calculated in \citet{planck_starlight} is also shown for reference, which predicts a lower microwave polarization intensity than observed in the Southern Hole for the vast majority of stars.
\newline\newline
This trend is not at all surprising, since most Heiles stars in the Southern Hole are not sensitive to the entire extent of the dust column along the line of sight. There has been analysis done showing that, at high galactic latitudes, most of the dust emission is due to a magnetized structure within 300 pc, which corresponds to the edge of the Local Bubble \citep{local_bubble}. However, the majority of the stars in Fig.~\ref{microwave_fig} are closer than 300 pc, and therefore they still might not be sensitive to the entire line of sight even if the analysis in \citet{local_bubble} holds true for patches at low galactic latitudes, like the Southern Hole. The dust behind those stars is likely aligned in a different direction due to the fluctuations in the magnetic field structure of the Milky Way, resulting in additional mixing of polarization directions and an overall reduction of the polarization fraction that can be seen in sub-millimetre measurements capturing the entire dust column, but not in starlight.
\newline\newline
This bias is mostly avoided in \citet{planck_starlight}, where additional filtering of the dataset is applied to ensure that the selected stars probe most of the corresponding dust columns. In this study, however, the number of available stars is extremely limited compared to \citet{planck_starlight} and the same filtering cannot be applied. Therefore, one may interpret the line of best fit in Fig.~\ref{microwave_fig} as a lower limit on the microwave polarization intensity rather than a mean value estimator.

\section{Discussion}
All of the analysis in this paper was carried out using the small amount of existing starlight polarization data in the Southern Hole. As seen in Fig.~\ref{patch}, there are only about 70 sources in the Southern Hole that currently have polarization measurements, which means only 0.175 stars per square degree in that region. This is far below the star density needed for high confidence analysis. Work outlined in \citet{roman} provides analysis of the number of sources needed for starlight reconstruction of dust templates as a function of the desired reconstruction error. The work suggests that polarization measurements for tens of thousands of stars are required to adequately proceed with the reconstruction of the dust template. This paper shows how polarization data and distance data can be combined to perform such reconstruction for the Southern Hole, but the small amount of stars per square degree that have existing polarization data cannot be ignored. We hope this will motivate PASIPHAE to make the Southern Hole its next target and deliver more uniform coverage of the region and precise polarization measurements along with it. If this work is carried out, it will create a great opportunity for the analysis of the Southern Hole using the methods in this paper, especially given that the Gaia DR2 catalog contains distance data for over 41,000 stars per square degree, and allow for the creation of a helpful template for CMB projects to understand the foreground systematics and subtract its effects from their B-mode searches.
\newpage

\section{Acknowledgements}
This work has made use of the VizieR database, operated at CDS, Strasbourg, France.
\newline
This work has made use of NASA's Astrophysics Data System.
\newline
This work has made use of data from the European Space Agency (ESA) mission {\it Gaia} (https://www.cosmos.esa.int/gaia), processed by the {\it Gaia} Data Processing and Analysis Consortium (DPAC, https://www.cosmos.esa.int/web/gaia/dpac/consortium). Funding for the DPAC has been provided by national institutions, in particular the institutions participating in the {\it Gaia} Multilateral Agreement.
\newline
The authors thank Gary Cole for his comments on a draft of this paper.
\newline
The authors thank Gina Panopoulou for her comments on a draft of this paper.
\newline
The authors thank Andrew Friedman and Grant Teply and our colleagues on POLARBEAR and BICEP teams for useful insights on this paper.

\bibliographystyle{ptapap}
\bibliography{pta4authors}

\end{document}